# SafeGuard: A Lightweight Client-Server Architecture for Real-Time Endpoint Threat Detection and Response

Francis Gideon Oghie[1], Shittu DivineDavid Abolanle[1]

[1]Department of Cyber Security Science, School of Information and Communication Technology, Federal University of Technology, Minna, Nigeria

**Abstract**

Endpoint devices, including laptops, smartphones, and tablets, remain among the most frequently targeted assets in organizational environments and are often implicated in security incidents and data breaches. While commercial Endpoint Detection and Response (EDR) platforms provide extensive monitoring and response capabilities, their deployment costs and operational complexity may limit adoption in small-to-medium enterprises, educational institutions, and other resource-constrained environments. This paper presents SafeGuard, a lightweight, three-tier client–server architecture, comprising an endpoint agent, a central server, and an administrative dashboard, for real-time endpoint monitoring, threat reporting, and administrative response. Rather than advancing threat-detection algorithms, the contribution of this work lies in the design, implementation, and validation of a low-cost, deployable endpoint security architecture.

The endpoint agent is developed with Flutter and extended with Kotlin for Android-level system access; the Node.js-based central server is responsible for authentication and secure message orchestration; and the administrative dashboard provides live endpoint visibility and remote response capabilities, including device locking, application removal, and warning notification dispatch. Threat identification is implemented through signature-based comparison against a maintained threat database, a deliberate design decision intended to minimize computational overhead while preserving practical monitoring functionality. Secure communication is achieved through encrypted WebSocket (WSS) channels, JSON Web Token (JWT)-based authentication, and HMAC message integrity verification.

The system was evaluated through unit, integration, and system-level testing, including a simulated 50-endpoint deployment and preliminary security validation against unauthorized access, replay attempts, and SQL injection attacks. Results showed an average command-dispatch latency of approximately 1.5 seconds, remaining below 2 seconds under simulated load. These findings indicate that an open-source technology stack can deliver real-time endpoint visibility and coordinated administrative response without the licensing costs of commercial EDR platforms. Limitations, including reliance on static threat signatures and Android-focused implementation depth, are discussed together with directions for future extension through advanced detection mechanisms and broader platform support.

**Keywords:** endpoint security, threat monitoring, client–server architecture, real-time response, Flutter, secure communication

## 1. Introduction

The proliferation of endpoint devices, including laptops, smartphones, and tablets, has reshaped how organisations communicate, share data, and conduct business, while simultaneously widening the attack surface available to adversaries. Endpoints are consistently identified as a leading vector in security incidents, a risk compounded by the adoption of Bring Your Own Device (BYOD) policies and remote work arrangements that introduce heterogeneous, often unmanaged devices into organisational networks [1], [2]. A single compromised endpoint can serve as a foothold for broader network intrusion, making endpoint visibility and rapid response a critical rather than peripheral component of organisational cybersecurity posture [3].

Commercial Endpoint Detection and Response (EDR) platforms address this need with comprehensive monitoring, behavioural analytics, and automated response capabilities [4]. However, these platforms are generally designed for, and priced toward, large enterprise deployments. Their licensing costs, infrastructure requirements, and operational complexity present a meaningful barrier to adoption for small-to-medium enterprises (SMEs), schools, and other resource-constrained organisations [5] — the same organisations that often lack dedicated security staff and are therefore least equipped to compensate for the absence of automated tooling. At the opposite end of the market, low-cost or open-source endpoint tools tend to operate as standalone antivirus utilities: they detect threats locally but provide administrators no consolidated view across endpoints and no mechanism for centralised, remote response. Many also rely on periodic polling rather than persistent, low-latency channels, delaying the interval between threat detection and administrative awareness, and are built with security analysts rather than IT generalists in mind, limiting their practical usability in organisations without dedicated security staff. The result is a gap between two extremes — full-featured but costly enterprise EDR, and inexpensive but centrally unmanaged antivirus tooling — with little practical middle ground for organisations that need real-time, centrally coordinated endpoint visibility but cannot justify enterprise licensing costs.

This paper addresses that gap by presenting SafeGuard, a lightweight client–server architecture for real-time endpoint monitoring, threat reporting, and administrative response, built entirely from open-source, no-license-cost components. The system comprises an endpoint agent developed with Flutter and extended with Kotlin for Android-level system access, a Node.js-based central server that authenticates devices and orchestrates secure message exchange over encrypted WebSocket connections, and an administrative dashboard providing live endpoint visibility and remote response actions such as device locking, application removal, and warning dispatch. Threat identification is implemented through signature-based comparison against a maintained threat database — a deliberate design choice that favours low computational overhead and straightforward auditability over the complexity of behavioural or machine-learning-based detection, and one that keeps the detection module decoupled from the surrounding communication and response architecture so that it could, in principle, be replaced by a more sophisticated backend without redesigning the system. The specific contributions of this paper are as follows:

- A validated, three-tier reference architecture demonstrating that real-time endpoint monitoring and centrally coordinated administrative response can be implemented using entirely open-source, no-license-cost components.
- An empirical characterisation of command-dispatch latency under simulated multi-endpoint load, providing a preliminary performance baseline for similarly scoped systems.
- A working, security-validated implementation, tested against unauthorised access, replay attempts, and SQL injection, demonstrating the architecture's practical resilience rather than only its theoretical design.

Consistent with these contributions, this paper's scope is architectural and empirical rather than algorithmic: it does not advance the state of the art in threat-detection methods, and the signature-based detection module used here is treated as a replaceable component rather than a research contribution in its own right. Similarly, while the endpoint agent's interface layer is built with a cross-platform framework, the system-level integrations evaluated in this study — device administration and native API access — were implemented and tested on Android only.

**Threat Model**

To clarify the security guarantees this work does and does not claim, we state explicitly the adversary model assumed throughout this paper. The threat model considered is a network-level and application-level adversary positioned to intercept or inject traffic between an endpoint and the central server, to attempt unauthorized access to the administrative dashboard or server API, to introduce a known-malicious application onto a monitored endpoint, or to attempt to replay previously captured messages to trigger unintended administrative actions. Within this model, SafeGuard is designed to: encrypt all client–server communication to prevent passive eavesdropping; reject requests that do not carry a valid, unexpired authentication token; detect applications matching entries in its threat signature database; and resist message replay through the integrity mechanisms described in Section 3.1.

This threat model explicitly excludes several adversary capabilities. It does not assume protection against a sufficiently privileged, physically present attacker who has already compromised the endpoint device at the operating-system or firmware level, since an attacker with such access could circumvent the agent entirely. It does not assume detection of threats absent from the signature database, including zero-day malware or novel attack techniques, consistent with the detection-scope limitation stated in Section 3.3. It does not assume resistance to a compromised or malicious administrator account, since the system's design delegates trust to authenticated administrators by intent. It also does not assume resilience against denial-of-service attacks targeting the central server's availability, or against social-engineering attacks aimed at obtaining administrator credentials, neither of which were evaluated in this study. Stating these exclusions explicitly is intended to make clear what the security validation reported in Section 4.3 does, and does not, establish.

The remainder of this paper is organised as follows. Section 2 reviews related work on endpoint security, secure communication protocols, and cross-platform development. Section 3 describes the materials and methods, including system architecture, technology justification, and testing methodology. Section 4 presents and discusses experimental results. Section 5 states the limitations of this study. Section 6 concludes the paper and outlines directions for future research.

## 2. Related Work

Endpoint security has evolved from isolated, signature-based antivirus software toward integrated platforms that combine secure communication, authentication, and continuous monitoring. Early endpoint protection, dating to the 1980s and 1990s, relied on signature matching against known malware databases and operated with minimal regard for centralized policy enforcement, as most endpoints functioned in relatively isolated network environments [6]. The proliferation of laptops and early internet connectivity in the 2000s exposed the limitations of this model, prompting the integration of intrusion detection systems and firewalls at the endpoint level [7]. The smartphone era of the 2010s, accompanied by the rise of Bring Your Own Device (BYOD) policies, further expanded the diversity of endpoints requiring protection and motivated the adoption of end-to-end encryption protocols such as TLS and multi-factor authentication mechanisms [8], [9]. These developments culminated in modern Endpoint Detection and Response (EDR) platforms, which combine real-time monitoring with automated response capabilities [10].

Despite their capability, EDR platforms introduce their own operational challenges. Continuous, high-fidelity monitoring generates substantial volumes of telemetry, and the resulting alert volume can overwhelm security teams, a phenomenon documented as alert fatigue that risks obscuring genuine threats amid low-priority notifications [11]. This usability burden is compounded by the resource intensity of comprehensive monitoring, which can degrade performance on constrained devices [12]. These characteristics are not incidental; they follow from a design orientation toward large security operations centers with dedicated analyst staff, rather than toward the IT generalists who administer security in smaller organizations. This motivates examining lighter-weight alternatives, even where such alternatives necessarily trade detection sophistication for operational simplicity.

Secure communication between endpoints and central infrastructure depends on standardized cryptographic protocols. Transport Layer Security (TLS) 1.3 remains the de facto standard for encrypting data in transit, addressing confidentiality and integrity through authenticated encryption and forward secrecy [13]. Authentication and authorization are typically layered on top of transport security;

the OAuth 2.0 framework, for instance, enables token-based authorization without requiring endpoints to expose long-lived credentials, reducing the impact of any single compromised token [14]. Message-level integrity is commonly provided through hash-based authentication codes such as HMAC, which detect tampering independent of the underlying transport channel [15]. Together, these mechanisms form a well-established toolkit for securing client–server communication; the design challenge for a lightweight endpoint system is less about inventing new cryptographic primitives and more about composing existing, well-vetted ones into an architecture that remains usable and performant on resource-constrained infrastructure.

A persistent challenge across this literature is the tension between platform-specific depth and cross-platform reach. Security-relevant operations such as device administration, secure key storage, and system-level telemetry are frequently exposed only through platform-native APIs, which constrains tools built on cross-platform frameworks to either reimplement these operations per platform or accept reduced functionality outside their primary target [16]. Flutter, an open-source UI toolkit that compiles to native machine code rather than relying on a JavaScript bridge or WebView, has gained adoption specifically because it allows a single codebase to deliver a consistent interface across Android, iOS, web, and desktop targets [16]. However, Flutter's plugin architecture still requires native code, typically via platform channels to Kotlin or Swift, for operations that touch device-level system APIs, meaning that "cross-platform" in practice describes the presentation layer more reliably than it describes deep system integration [16].

Taken together, the literature suggests three observations relevant to this study. First, the security and usability characteristics of full-featured EDR platforms are well documented but are calibrated for enterprise contexts with dedicated security staff, leaving a gap for organizations that need centralized visibility without that operational overhead. Second, the cryptographic and authentication building blocks required for secure endpoint communication are mature and standardized, such that a new system's contribution lies in their composition and deployment rather than in novel protocol design. Third, claims of cross-platform endpoint security coverage should be read carefully, since the underlying system-level integrations on which security actions depend are frequently platform-specific in practice regardless of the UI framework used. This study positions itself accordingly: it composes established protocols (TLS/WSS, token-based authentication, HMAC integrity) into a three-tier architecture aimed at the underserved middle ground between standalone antivirus tools and enterprise EDR, while being explicit that its system-level implementation and validation are scoped to Android rather than claimed across all platforms the UI framework nominally supports.

## 3. Materials and Methods

### 3.1 System Architecture

SafeGuard follows a three-tier client–server architecture comprising an endpoint agent, a central server, and an administrative dashboard. This separation ensures that endpoints never communicate directly with one another, reducing the risk of peer-to-peer infection spread and concentrating monitoring and policy enforcement at a single, auditable point. The endpoint agent is installed on each monitored device and is responsible for registering with the server using a unique device identifier, periodically reporting system metrics (CPU usage, installed applications), comparing installed applications against a locally cached threat signature database, and listening for administrative commands over a persistent connection. The central server authenticates both endpoints and administrators, maintains WebSocket connections for bidirectional real-time communication, and relays administrator-issued commands to the appropriate endpoint. Data storage follows a hybrid model: Firebase (Realtime Database/Firestore) handles real-time state synchronization, push notifications, and authentication support, while a separate relational database (SQL) stores structured logs, threat reports, and command history under a third normal form (3NF) schema, chosen to minimize data redundancy in records that are queried and audited after the fact. The administrative dashboard provides a live view of connected devices and their status, displays threat alerts as they are reported, and allows the administrator to issue remote response actions, including locking a device, uninstalling a flagged application, or sending a warning notification.

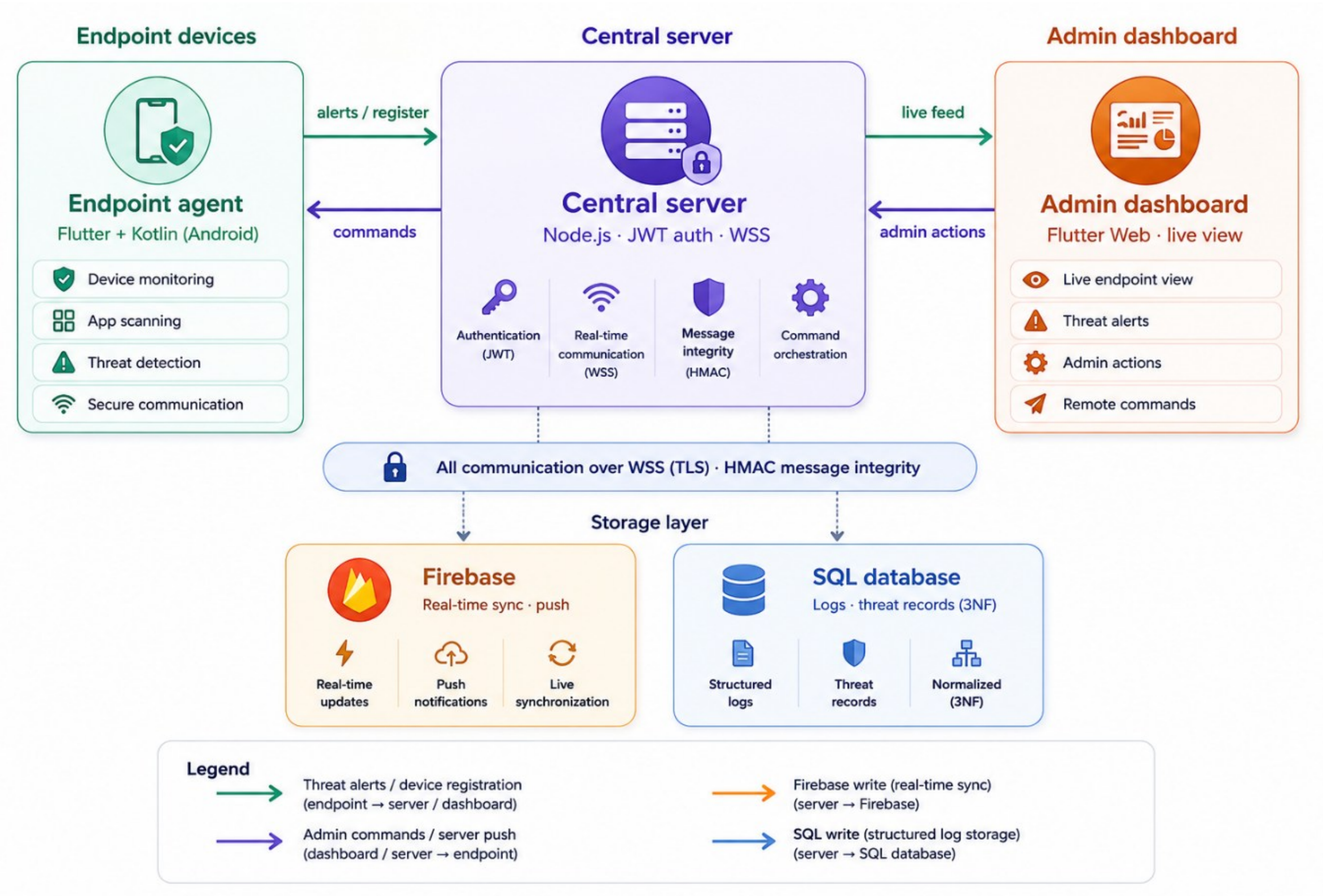

Figure 1. SafeGuard three-tier system architecture

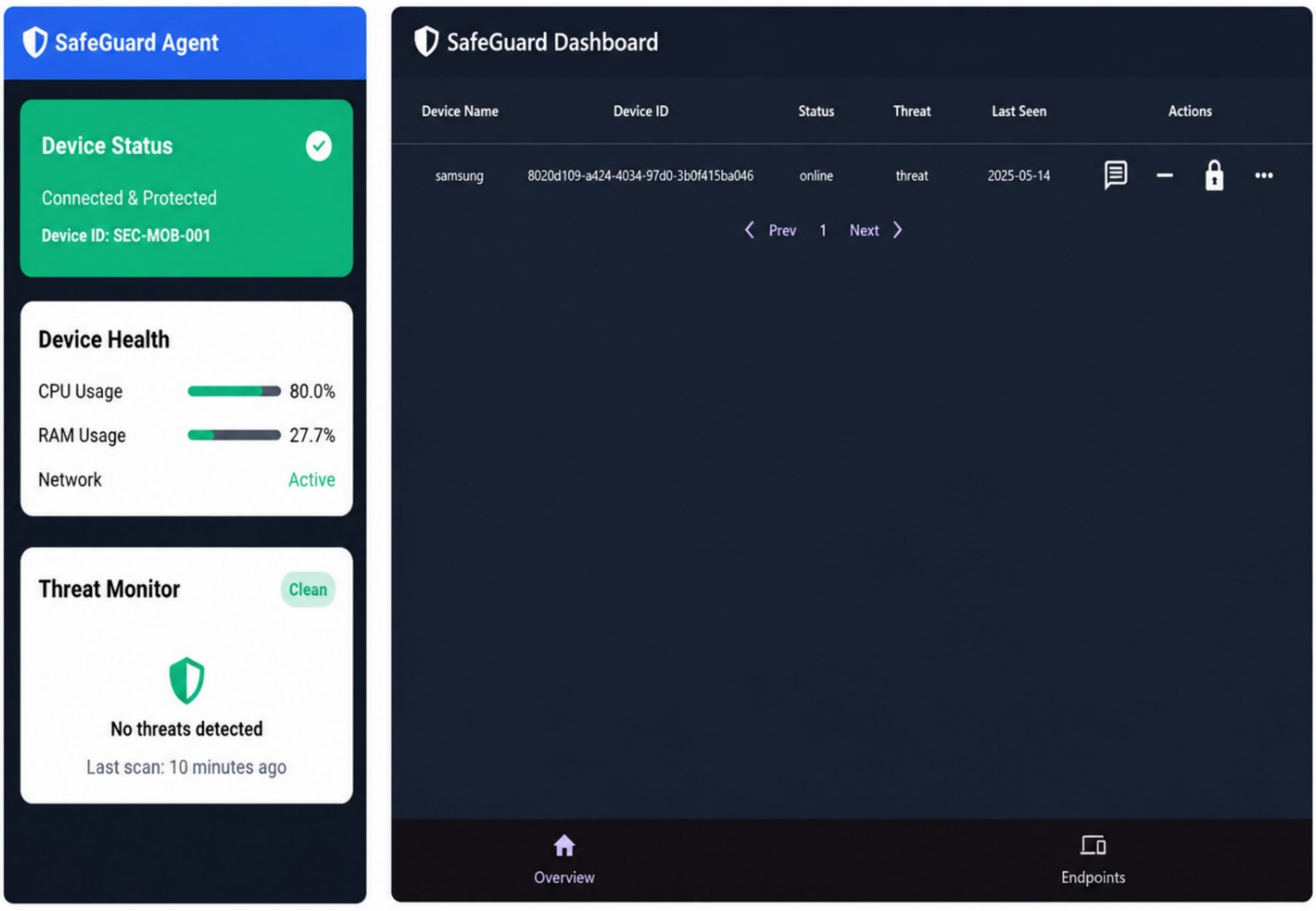

Figure 2. SafeGuard interface overview: (a) endpoint agent device status screen (b) administrative dashboard device list

Communication between tiers occurs over WebSocket connections secured with TLS (wss://), avoiding the latency overhead of repeated HTTP polling and enabling the server to push commands to endpoints immediately rather than waiting for the next scheduled check-in. Authentication is handled through JSON Web Tokens (JWT) issued at login and validated on each privileged request, and message integrity is reinforced with HMAC signing, consistent with established practice for keyed message authentication [13]. This combination addresses, within the scope of this system, the absence of end-to-end security and centralized policy enforcement identified as a recurring weakness of lower-cost endpoint tools.

### 3.2 Technology Selection and Justification

The endpoint agent's user interface is implemented in Flutter, an open-source toolkit that compiles to native machine code and renders through its own graphics engine rather than relying on platform WebViews, allowing a single codebase to target Android, iOS, web, and desktop with a consistent interface [14]. Flutter was selected over alternatives such as React Native primarily for this rendering approach and its growing adoption in both startup and enterprise contexts. However, as established in Section 2, Flutter's plugin architecture does not provide direct access to platform-level system APIs; operations such as querying installed applications, locking a device, or uninstalling an application require native platform code. Kotlin was therefore used, via Flutter's platform channel mechanism, to implement these Android-specific system-level operations. This pairing reflects a deliberate division of labor: Flutter supplies a maintainable, consistent presentation layer, while Kotlin supplies the system access that the presentation layer cannot reach on its own. The central server is implemented in Node.js, chosen for its native support of asynchronous, event-driven I/O, which suits the persistent-connection, message-passing pattern WebSocket communication requires, and for its operational simplicity relative to heavier enterprise backend stacks.

### 3.3 Threat Detection Approach

Three broad detection strategies are established in the literature: signature-based detection, which compares observed artifacts against a database of known threat indicators and offers high accuracy against known threats but no protection against novel ones; anomaly-based detection, which flags deviations from a learned baseline of normal behavior and can catch unknown threats at the cost of higher false-positive rates and the overhead of establishing reliable baselines; and heuristic-based detection, which applies rule-based pattern matching to balance the two [15], [16], [17].

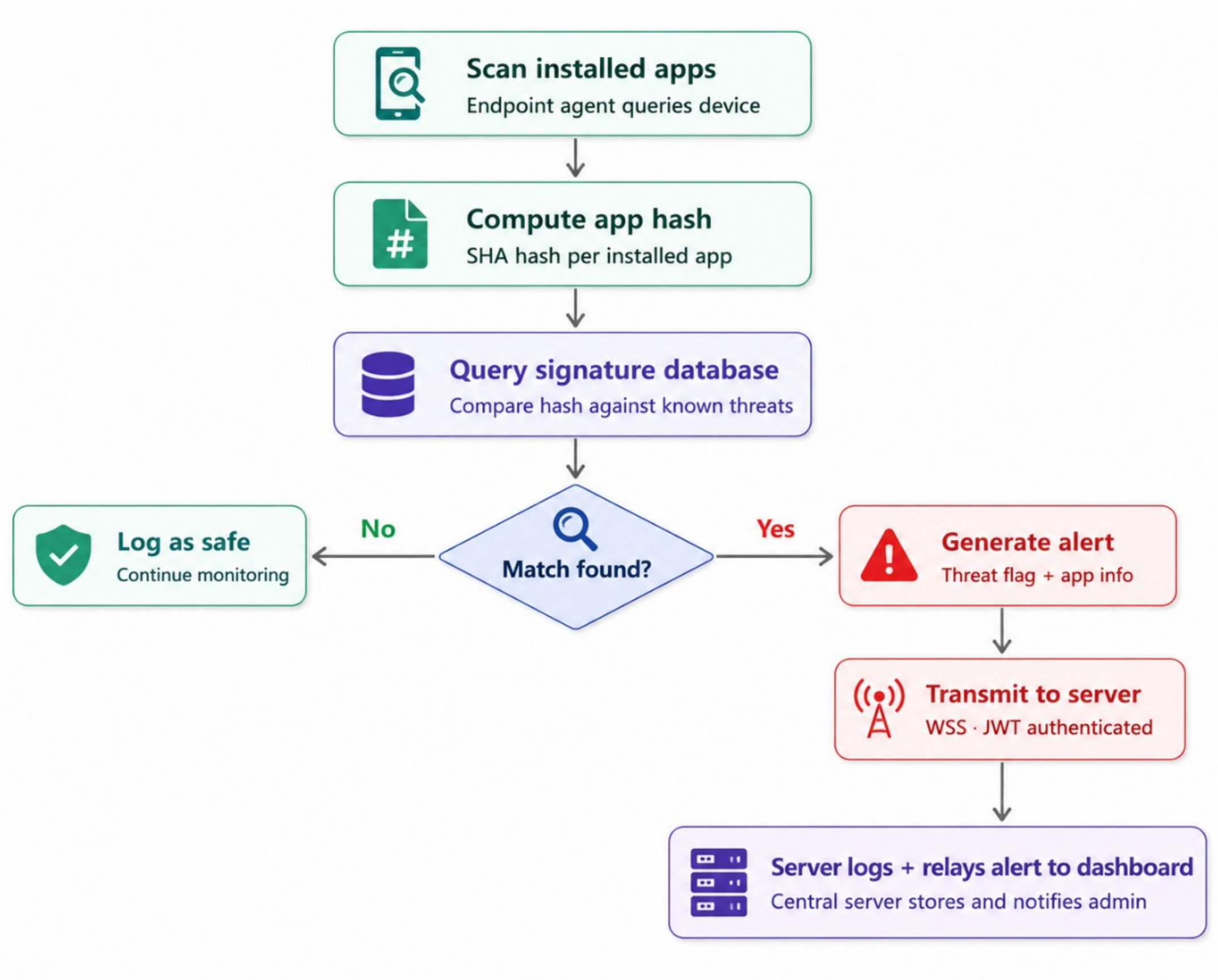

Figure 3. Signature-based threat detection workflow

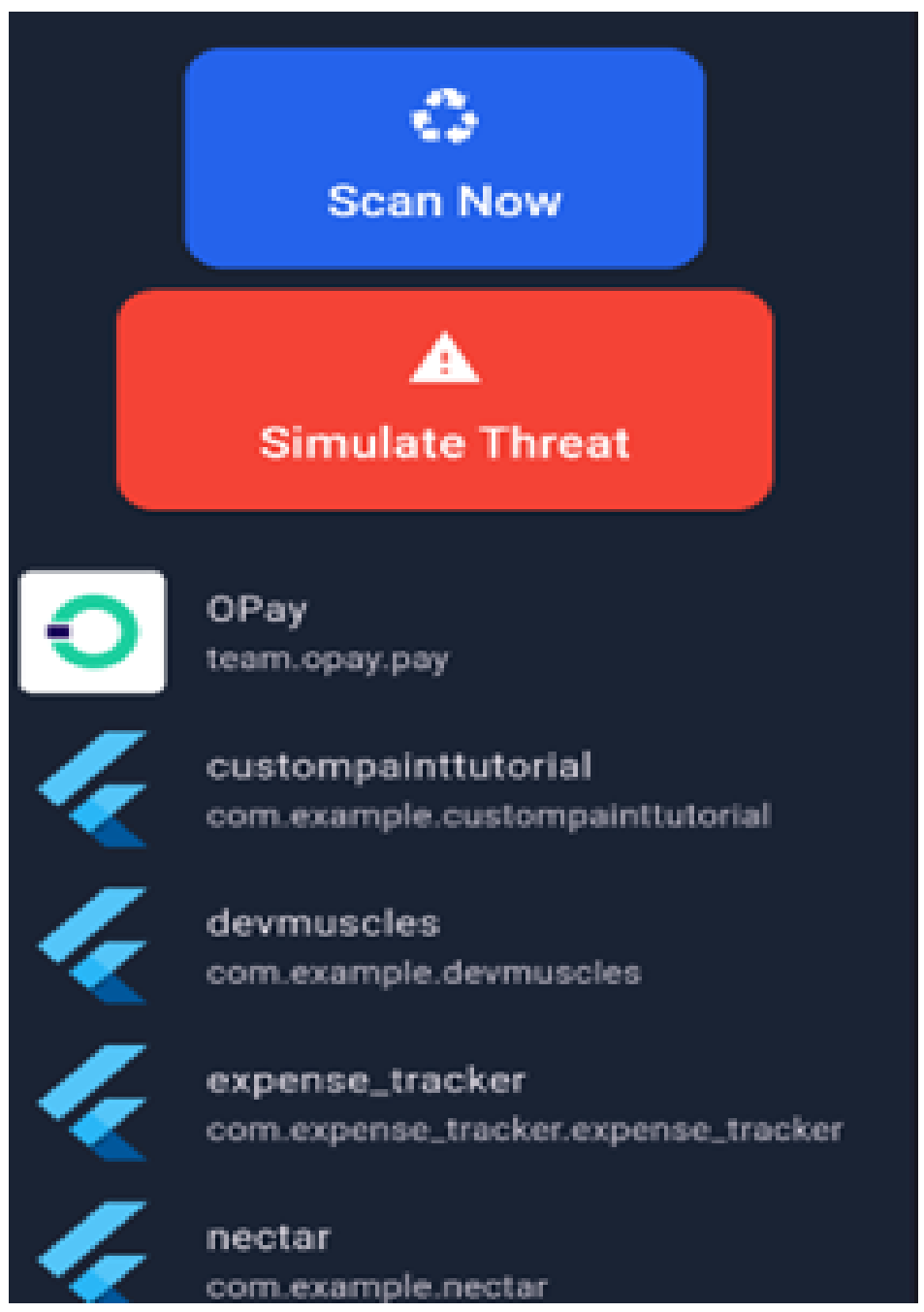

Figure 4. Endpoint agent application scanning interface

SafeGuard implements signature-based detection: the endpoint agent computes a hash of each installed application and compares it against a maintained threat signature database, reporting a match as a potential threat. This choice was made deliberately rather than by default. Signature comparison has a low, predictable computational cost, requires no training data or baseline-establishment period, and produces auditable, explainable detection decisions, properties that matter for a system intended to run continuously on resource-constrained, non-enterprise hardware. The acknowledged cost of this choice is that the system cannot detect previously unseen (zero-day) threats absent from its signature database; this limitation is revisited in Section 5. The detection module is implemented as a discrete component that receives application metadata and returns a match decision, a separation intended to allow the signature-based engine to be replaced by an anomaly-based or heuristic engine in future work without requiring changes to the surrounding communication and response architecture.

**3.4 Functional and Non-Functional Requirements**

Table 1 and Table 2 summarize the functional and non-functional requirements that guided system design and against which the implementation was evaluated.

***Table 1. Functional Requirements***

| ID | Requirement |
|---|---|
| FR1 | The administrator shall be able to view a list of connected endpoints. |
| FR2 | The endpoint agent shall continuously monitor installed applications using signature-based comparison. |
| FR3 | The system shall send threat alerts to the administrator in real time upon detection. |
| FR4 | The administrator shall be able to remotely issue actions (lock device, terminate application) on affected endpoints. |
| FR5 | All communications shall be encrypted using WebSocket over TLS. |
| FR6 | The system shall persist logs and threat reports for all endpoints. |

***Table 2. Non-Functional Requirements***

| ID | Requirement |
|---|---|
| NFR1 | The system shall support Android 8.0 and later. |
| NFR2 | The system shall target message round-trip latency below 500 ms. |
| NFR3 | The endpoint agent shall maintain minimal CPU and memory consumption. |
| NFR4 | The administrative interface shall remain responsive across screen sizes. |
| NFR5 | The backend shall be designed to scale toward larger endpoint counts. |

NFR1 makes explicit a scope decision carried throughout this study: the system's non-functional requirements were defined, implemented, and tested against Android, and claims of broader platform support are confined to the agent's interface layer rather than its full functional behavior.

**3.5 Development Environment**

The system was developed using a hybrid toolchain combining cross-platform and native Android development tools. Development was carried out on a Windows 10 machine (Intel Core i5, 8 GB RAM) using Visual Studio Code and Android Studio as the primary IDEs. The endpoint agent was built with Flutter SDK 3.x and tested on a physical Android device (Redmi Note 12, Android 13) and an Android Virtual Device (Pixel 5 emulator profile, API level 34). The central server was implemented in Node.js, with Firebase used for real-time synchronization and notification delivery, and SQLite/MySQL used for structured relational storage of logs and threat records. Postman was used for REST API testing, and Wireshark was used to inspect network traffic during security testing. Source code was managed using a GitHub repository.

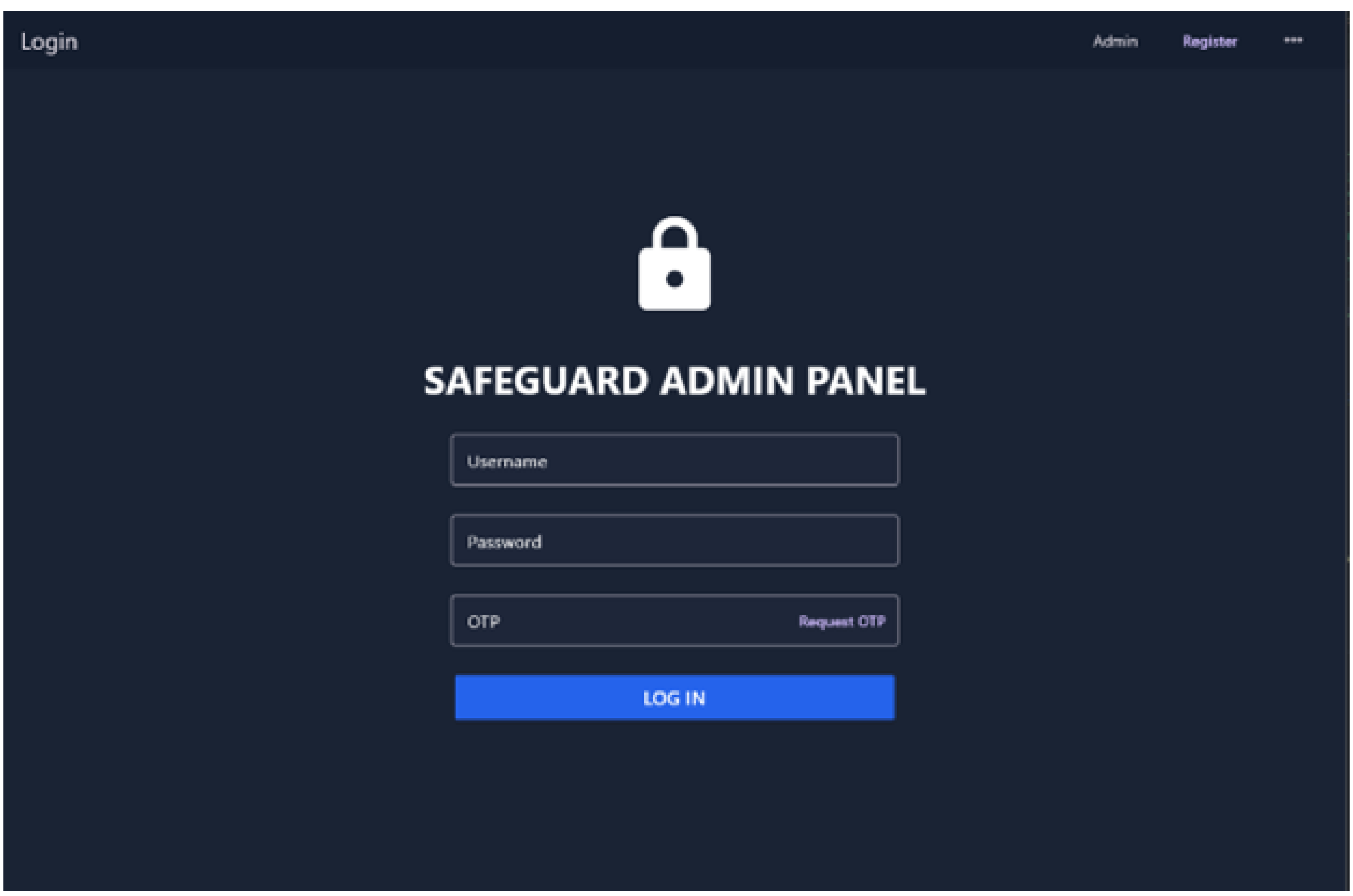

Figure 5. SafeGuard administrative dashboard login screen,

**3.6 Testing Methodology**

The system was evaluated through four stages of testing, each targeting a different level of system behaviour:

*Unit testing* verified individual module behavior in isolation: endpoint registration against the expected server response, threat detection against a known-malicious application hash, JWT validation rejecting tampered or expired tokens, and command dispatch routing a command to the correct endpoint identifier.

*Integration testing* verified behavior across module boundaries: that an endpoint registering with the server was correctly reflected in the dashboard's device list, that a threat alert generated by an endpoint was correctly logged by the server and displayed on the dashboard, and that an administrator-issued uninstall command was correctly dispatched by the server and executed by the target endpoint.

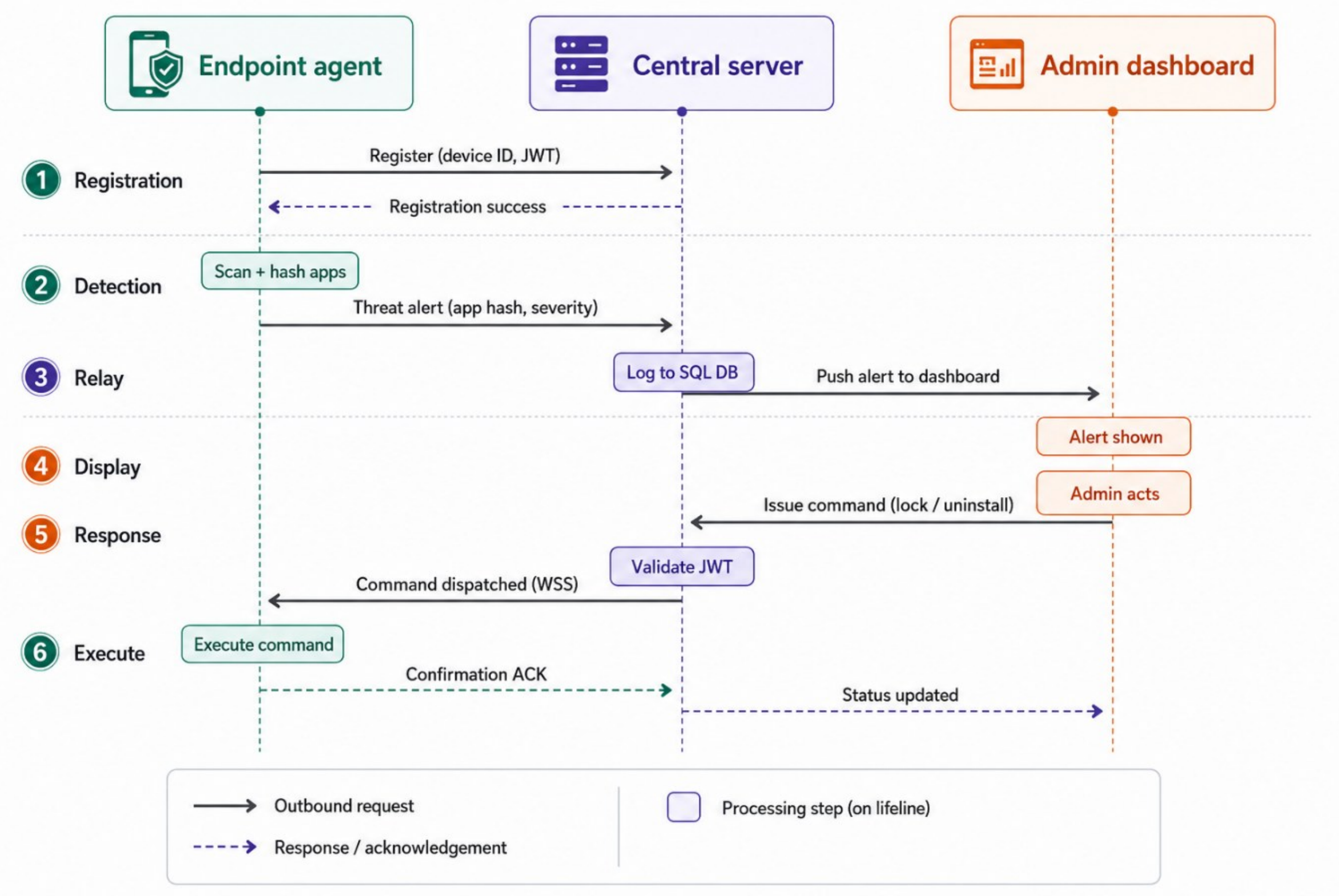

Figure 6. End-to-end threat response sequence.

*System testing* evaluated the assembled system under three scenarios: normal operation with multiple endpoints connected and reporting status; a simulated attack in which a known-malicious application signature was introduced to a test endpoint, triggering detection, alerting, and a subsequent administrator lock command; and a load scenario in which 50 simulated endpoint connections were established concurrently to observe whether response latency degraded under multi-endpoint load. We report this as a single load scenario rather than a multi-trial benchmark; the absence of repeated trials and statistical reporting (mean, variance, confidence intervals) is acknowledged as a methodological limitation in Section 5, and the resulting figures should be read as a preliminary observation rather than a statistically validated performance characterization.

*Security validation* assessed three specific attack patterns rather than constituting a comprehensive penetration test: confirmation that all client–server traffic remained encrypted under inspection with Wireshark; confirmation that requests bearing invalid or tampered JWTs were rejected by the server; and manual attempts at SQL injection against server input fields and at replaying captured WebSocket messages, both of which were unsuccessful against the implemented defenses. No automated security scanning tool, formal threat model, or third-party assessment was used, and these results should be read as preliminary validation of specific, named defenses rather than a general security audit. This scope is stated explicitly here, and is revisited in the Limitations section, to avoid overstating what was tested.

## 4. Results and Discussion

### 4.1 Functional Correctness

Unit and integration testing confirmed that the implemented system behaved as specified across all functional requirements listed in Table 1. Endpoint registration correctly returned a success response from the server and was reflected in the administrative dashboard's device list. Signature-based detection correctly flagged a known-malicious application hash and generated a threat alert that propagated from the endpoint agent to the server log and, in turn, to the dashboard. JWT validation correctly rejected requests bearing tampered or expired tokens, and administrator-issued commands (lock, uninstall, warning) were correctly routed to the intended endpoint and executed.

```
I/flutter ( 4153): [ThreatScan] Checking 22 hashes for threats...
I/flutter ( 4153): [ThreatScan] Hash: 021575009de92bce4322fa2db08fedb9 FOUND in DB. Status: Safe
I/flutter ( 4153): [ThreatScan] Hash: 908baabb26daa1d024c496916507ac46 FOUND in DB. Status: Safe
I/flutter ( 4153): [ThreatScan] Hash: 3c1f406dbb1c756d3755d545170e174e FOUND in DB. Status: Safe
I/flutter ( 4153): [ThreatScan] Hash: 3b49b8f0f0a57daef6bf574d7e66473b FOUND in DB. Status: Safe
I/flutter ( 4153): [ThreatScan] Hash: 477eafd43a581914a1127d08f067e658 FOUND in DB. Status: Safe
I/flutter ( 4153): [ThreatScan] Hash: ba23e908e16bb1cb6bc41308afea6b94 FOUND in DB. Status: Safe
I/flutter ( 4153): [ThreatScan] Hash: 1cf542583aa13b6ccfbf78eedb1372df FOUND in DB. Status: Safe
I/flutter ( 4153): [ThreatScan] Hash: 0534c918050551641e99c9f7ec760c46 FOUND in DB. Status: Safe
I/flutter ( 4153): [ThreatScan] Hash: f9817547c8f343564c63282cc455e1e1 FOUND in DB. Status: Safe
I/flutter ( 4153): [ThreatScan] Hash: 1cd8605bf0fcddf2af2f424e12ff164e FOUND in DB. Status: Safe
I/flutter ( 4153): [ThreatScan] Hash: 6930499e35a8960c471b20878f5a56ca FOUND in DB. Status: Safe
I/flutter ( 4153): [ThreatScan] Hash: 3ce63f2e1c320876f20292c2342c3bff FOUND in DB. Status: Safe
I/flutter ( 4153): [ThreatScan] Hash: b8b0a1c1405f2e862f0feabfc7be5912 FOUND in DB. Status: Safe
I/flutter ( 4153): [ThreatScan] Hash: d2476a176c567bc6e1763c801b2cc4d1 FOUND in DB. Status: Safe
I/flutter ( 4153): [ThreatScan] Hash: 64b8d942cc64bde0ee6f3bcf40cae38f FOUND in DB. Status: Safe
I/flutter ( 4153): [ThreatScan] Hash: 092539a30aa675f30d207453a1a7fe15 FOUND in DB. Status: Safe
I/flutter ( 4153): [ThreatScan] Hash: 214cb81d438fed44a3705ccaa5825dd9 FOUND in DB. Status: Safe
I/flutter ( 4153): [ThreatScan] Hash: 4e15969b0de06ada36cbed7a5f8c2407 FOUND in DB. Status: Safe
I/flutter ( 4153): [ThreatScan] Hash: 3fe250b4b7851853b59f868f862a0613 FOUND in DB. Status: Safe
I/flutter ( 4153): [ThreatScan] Hash: 3bc3c6a31ec8d6d03761ab755e7c5c97 FOUND in DB. Status: Safe
I/flutter ( 4153): [ThreatScan] Hash: d663f8367dfb4784d1faf66d5ecc0026 FOUND in DB. Status: Safe
I/flutter ( 4153): [ThreatScan] Hash: a80772177d5836b8c54e8b3d9030cea1 FOUND in DB. Status: Safe
I/flutter ( 4153): [ThreatScan] No threats detected.
I/flutter ( 4153): [ThreatScan] Scan complete.
```

Figure 7. Console output from the endpoint agent during signature-based threat scanning

These outcomes indicate that the three-tier architecture functions as designed for the scenarios tested: a registered endpoint, an inserted threat signature, and a single administrator issuing commands. They do not, on their own, establish how the system behaves under conditions outside these scenarios, such as concurrent administrator sessions, partial network failure, or endpoints rejoining after disconnection, none of which were exercised in testing.

### 4.2 Response Latency

Command-dispatch latency, measured as the interval between an administrator issuing a command and the endpoint executing it, averaged approximately 1.5 seconds during system testing. Under the 50-simulated-endpoint load scenario, latency remained below 2 seconds. Both figures exceed the 500 ms round-trip target set out as a non-functional requirement (NFR2) during system design; we report this gap directly rather than omit it, since the original target provides a useful reference point even though the implemented system did not meet it. The observed latency likely reflects the combined overhead of WebSocket message serialization, server-side authentication checks on each command, and database logging performed before relaying a command to its target endpoint, though this study did not isolate the contribution of each stage and so cannot attribute the gap to a specific cause with confidence.
These latency figures should be read as a single observation rather than a statistically characterized result. Testing was not repeated across multiple independent trials, and no variance, confidence interval, or distribution of latency values was recorded; only a single reported average is available for both the unloaded and 50-endpoint conditions. Likewise, no comparison was made against a baseline system, whether a competing open-source tool or a simplified version of SafeGuard itself, so the practical significance of "1.5 seconds" can only be assessed qualitatively. For the use case this paper targets, real-time visibility and response orchestration for SMEs and schools rather than sub-second automated containment in a security operations center, a latency in the low single-digit seconds is plausibly adequate for human-in-the-loop administrative response, since the administrator must still read the alert and decide on an action before any command is issued. Whether 1.5 seconds is fast enough for automated, unattended response scenarios is a separate question this study does not answer.

**4.3 Security Validation**

Communication traffic inspected with Wireshark during testing remained encrypted, consistent with the system's use of WebSocket connections over TLS. Requests using invalid or tampered JWTs were rejected by the server, and manual attempts at SQL injection against server input fields and at replaying captured WebSocket messages did not succeed against the implemented defenses. These results indicate that the specific defenses evaluated, transport encryption, token validation, and basic injection and replay resistance, functioned as intended in the tested scenarios.

***Table 3. Outcomes of evaluated attack patterns***

| **Attack pattern** | **Method** | **Outcome** |
|---|---|---|
| Unauthorized access (invalid/tampered JWT) | Manual request with invalid or expired token | Rejected by server |
| SQL injection | Manual injection attempts against server input fields | Unsuccessful |
| Message replay | Manual replay of captured WebSocket messages | Unsuccessful |

These outcomes reflect the three attack patterns specifically tested, conducted manually as described in Section 3.6, and should not be extrapolated to attack classes outside this table, including man-in-the-middle attacks against the TLS handshake, denial-of-service, or credential-based social engineering, none of which were evaluated in this study.

This should not be read as evidence of comprehensive security, and we are explicit about that boundary here rather than leaving it implicit. The injection and replay attempts were conducted manually, without an automated scanning tool, a documented payload set, or a formal threat model, and no third-party or adversarial review was performed. The result is better described as confirmation that a small number of specific, named attack patterns did not succeed, not as a general security assurance about the system. A more rigorous security evaluation, using established tools (e.g., OWASP ZAP, Burp Suite) and a documented test plan, is identified as a direction for future work in Section 6.

### 4.4 Comparative Positioning

Table 4 situates SafeGuard's documented feature set against the general capabilities of commercial EDR platforms, as characterized in Section 2 [4], [10]. This is a qualitative, literature-grounded comparison rather than a head-to-head empirical benchmark; no commercial EDR platform was deployed or tested alongside SafeGuard in this study, and the comparison should be read accordingly, as a positioning of design intent and documented capability rather than a measured performance comparison.

***Table 4. Qualitative comparison of SafeGuard and commercial EDR platforms***

| Feature | SafeGuard | Commercial EDR |
|---|---|---|
| Real-time endpoint monitoring | Yes | Yes |
| Centralized administrative dashboard | Yes | Yes |
| Remote response actions (lock, uninstall, alert) | Yes | Yes |
| Detection method | Signature-based only | Signature-based, heuristic, and behavioral/AI-based [4] |
| Licensing cost | None (open-source stack) | Typically subscription-based, scaled to endpoint count [4] |
| Platform coverage (system-level integration) | Android only (tested) | Typically Windows, macOS, and Linux, with mobile support varying by vendor [4] |
| Designed operator | IT generalist | Dedicated security analyst / SOC [10] |
| Zero-day detection capability | No | Varies; behavioral/AI-based platforms claim partial coverage [4] |

The comparison highlights that SafeGuard's value proposition is not feature parity with commercial EDR, but coverage of a narrower, specific need: centrally coordinated visibility and response at no licensing cost, for organizations where commercial EDR's cost or operational complexity is the primary barrier rather than its detection sophistication. Where detection sophistication is the priority, for example in environments facing targeted or novel threats, commercial EDR's broader detection methods remain the more appropriate choice.

### 4.5 Discussion

Read together, these results support a narrow but specific claim: the three-tier architecture described in Section 3 is functionally correct for its core registration, detection, alerting, and command-dispatch workflows, and remains operational under a simulated 50-endpoint load without the tested defenses failing. This is consistent with the architectural literature reviewed in Section 2, which establishes that TLS/WSS, JWT, and HMAC are well-vetted mechanisms individually; what this study adds is evidence that composing them into a lightweight, three-tier system, built without commercial licensing costs, produces a working system rather than only a theoretical design.

The results do not support broader claims this study does not make. They do not establish that SafeGuard detects threats as effectively as a commercial EDR platform, since detection here is limited to signature matching against a maintained database and was not evaluated for detection rate, false-positive rate, or performance against unknown threats. They do not establish performance at a scale beyond 50 simulated endpoints, nor performance on real, heterogeneous hardware rather than a single test device and emulator. They do not establish security robustness beyond the three specific attack patterns tested. Within these boundaries, the results are consistent with the paper's stated contribution: a validated architecture and a preliminary performance and security baseline, rather than an advance in detection methodology or a production-ready security guarantee.

## 5. Limitations

This study has several limitations, grouped here by category rather than presented as an undifferentiated list, since they carry different implications for how the results should be interpreted.

**Detection scope.** SafeGuard's threat identification relies on signature-based comparison against a maintained database of known threat indicators. As established in Section 3.3, this was a deliberate design choice favoring predictable computational cost and explainability over broader coverage, but its direct consequence is that the system cannot detect previously unseen, zero-day threats, or any threat absent from its signature database at the time of comparison. This is not a defect to be fixed within the present architecture; it is the boundary of what a signature-based engine can do, and any future extension toward anomaly-based or heuristic detection (see Section 6) would need to be evaluated on its own terms.

**Platform scope.** The system's interface layer is built with a cross-platform framework, but its system-level integrations, including device locking and application removal, were implemented and tested on Android only, using a single physical device and one emulator profile. No implementation or testing was performed on iOS or desktop platforms, despite these being nominally reachable by the underlying UI framework. Claims of cross-platform support in this paper are therefore limited to the presentation layer and should not be read as evidence of equivalent functional depth across platforms.

**Evaluation rigor.** The performance results reported in Section 4 are drawn from a single test run rather than repeated independent trials, and no variance or confidence interval accompanies the reported latency figures. No baseline system, whether a competing open-source tool or an internal ablation of SafeGuard's own components, was used for comparison, which limits the latency results to a standalone observation rather than a comparative performance claim. Similarly, the 50-endpoint stress test used simulated rather than physically distinct endpoint devices, which may understate resource contention effects (e.g., shared network conditions, device-level CPU throttling) that would arise from genuinely heterogeneous hardware operating concurrently.

**Security validation depth.** The security checks reported in Section 4.3 were conducted manually against three specific attack patterns (unauthorized access via invalid tokens, SQL injection, and message replay), without an automated scanning tool, a documented and reproducible payload set, a formal threat model, or independent third-party review. These results confirm that the named defenses held against the specific attempts made; they do not constitute a general security audit, and the system's resistance to attack classes not tested here, including denial-of-service, man-in-the-middle attacks against the TLS handshake itself, or social-engineering-based compromise of administrator credentials, remains unevaluated.

**Feature completeness.** Several features identified during system design, including remote data wipe and network-level traffic monitoring, were not implemented, primarily due to platform permission restrictions encountered during development and the resource constraints of the development environment. Their absence means the response actions evaluated in this study (device lock, application removal, warning dispatch) represent a subset of the response capabilities a deployed system of this kind would likely need.

**Deployment scale.** All testing was conducted in a controlled development and laboratory setting. The system has not been deployed or evaluated in a production organizational environment, has not been tested at the scale of hundreds or thousands of endpoints, and has not been exposed to real, uncontrolled network conditions or genuine adversarial activity over an extended period. The results reported here should accordingly be read as evidence of feasibility under controlled conditions rather than evidence of production readiness.

Taken together, these limitations are consistent with the scope established at the outset of this paper: SafeGuard is presented as a validated architecture and a preliminary empirical baseline, not as a finished, enterprise-ready product or a contribution to detection science. The directions for addressing several of these limitations are discussed in Section 6.

**6. Conclusion and Future Work**

This paper presented SafeGuard, a lightweight, three-tier client–server architecture for real-time endpoint monitoring, threat reporting, and administrative response, built entirely from open-source, no-license-cost components. The system was designed around a clear gap identified in Sections 1 and 2: the absence of a practical middle ground between costly, operationally complex commercial EDR platforms and low-cost antivirus tools that offer no centralized visibility or coordinated response. The endpoint agent, central server, and administrative dashboard were implemented and evaluated through unit, integration, and system-level testing, including a simulated 50-endpoint load scenario and preliminary security validation against unauthorized access, SQL injection, and replay attempts. The system functioned correctly across the scenarios tested, with command-dispatch latency averaging approximately 1.5 seconds and remaining below 2 seconds under simulated load.

Consistent with the scope set out from the outset, this work's contribution is architectural and empirical: it demonstrates that real-time, centrally coordinated endpoint monitoring and response can be built and validated using freely available, open-source tooling, rather than advancing the state of the art in threat-detection algorithms. The signature-based detection module used here was treated throughout as a replaceable component, and the results reported in Section 4, together with the limitations stated plainly in Section 5, are intended to be read as a preliminary baseline rather than a finished or production-ready system.

Several directions for future work follow directly from the limitations identified in Section 5. Addressing the detection scope limitation would involve incorporating anomaly-based or machine-learning-based detection, for example using behavioral models trained on endpoint telemetry such as CPU usage patterns or network traffic anomalies, to complement signature matching and provide some capability against previously unseen threats; given the architecture's decoupling of the detection module from the surrounding communication layer, this could in principle be introduced without redesigning the rest of the system. Addressing the static-signature limitation more narrowly would involve integrating external threat intelligence feeds (e.g., VirusTotal, AlienVault OTX) so the signature database can be updated dynamically rather than maintained manually. Addressing the platform-scope limitation would require extending implementation and testing to Windows, macOS, Linux, and iOS endpoints, using platform-specific APIs or a multiplatform toolkit to provide the same depth of system-level access on each platform that this study achieved on Android alone; this is necessary before any claim of genuine cross-platform parity, rather than cross-platform UI presentation, can be made. Addressing the deployment-scale limitation would involve testing on physically distributed, heterogeneous hardware at endpoint counts well beyond the 50 simulated in this study, deploying the server on cloud infrastructure with load balancing and clustering, and reporting performance results across repeated trials with appropriate statistical treatment rather than single-run observations. Addressing the security-validation limitation would involve a structured security assessment using established tools and a documented, reproducible test methodology, ideally including independent or third-party review. Finally, extending the system's administrative capabilities, for example with remote data wipe, network isolation, forensic log export, and alignment with standards such as ISO 27001 or GDPR, would move the system meaningfully closer to a deployable tool for the resource-constrained organizations it is intended to serve.

**References**

[1] Verizon. (2024). *2024 data breach investigations report*. Verizon Business. Retrieved June 2026, from https://www.verizon.com/business/resources/reports/dbir/

[2] Scarfone, K., & Souppaya, M. (2019). *Guide to enterprise telework, remote access, and bring your own device (BYOD) security* (NIST Special Publication 800-46 Rev. 2). National Institute of Standards and Technology. https://doi.org/10.6028/NIST.SP.800-46r2

[3] Buecker, A., Campos, A., Cutler, P., Hu, A., Jeremiah, G., Matsui, T., & Zarakowski, M. (2012). *Endpoint security and compliance management design guide using IBM Tivoli Endpoint Manager* (IBM Redbooks, SG24-7980-00). IBM Corporation. https://www.redbooks.ibm.com/redbooks/pdfs/sg247980.pdf

[4] Gartner. (2024). *Magic quadrant for endpoint protection platforms*. Gartner, Inc.

[5] Waterson, D. (2020, October 28). Manage endpoints to protect the weakest link in the security chain. *Dave Waterson on Security*. https://dwaterson.com/2020/10/28/manage-endpoints-to-protect-the-weakest-link-in-the-security-chain/

[6] Melih, O. (2025, May 31). The evolution of endpoint security: From antivirus to unified zero trust. *Melih*. https://melih.com/the-evolution-of-endpoint-security-from-antivirus-to-unified-zero-trust/

[7] Padhiar, S., & Patel, R. (2023). Outside the closed world: On using machine learning for network intrusion detection. In *Proceedings of the International Conference on Information and Communication Technology for Intelligent Systems* (pp. 265–270). Springer Nature Singapore. https://doi.org/10.1007/978-981-19-9221-5_22

[8] Rescorla, E. (2018). *The Transport Layer Security (TLS) protocol version 1.3* (RFC 8446). Internet Engineering Task Force. https://doi.org/10.17487/RFC8446

[9] National Institute of Standards and Technology. (2017). *Digital identity guidelines: Authentication and lifecycle management* (NIST Special Publication 800-63B). U.S. Department of Commerce. https://doi.org/10.6028/NIST.SP.800-63b

[10] Holzinger, A., Saranti, A., & Angerschmidt, P. (2025). Managing data overload in EDR systems: A usability perspective. *Software: Practice and Experience, 55*(2), 234–248. https://doi.org/10.1002/spe.3123

[11] Bhadauria, S., & Sanyal, S. (2024). Resource-efficient EDR for IoT devices. *Journal of Cybersecurity, 12*(1), 112–125. https://doi.org/10.1093/cybsec/tyaa001

[12] Hardt, D. (Ed.). (2012). *The OAuth 2.0 authorization framework* (RFC 6749). Internet Engineering Task Force. https://doi.org/10.17487/RFC6749

[13] Krawczyk, H., Bellare, M., & Canetti, R. (1997). *HMAC: Keyed-hashing for message authentication* (RFC 2104). Internet Engineering Task Force. https://doi.org/10.17487/RFC2104

[14] Google. (2024). *Flutter architectural overview*. Flutter Documentation. Retrieved June 2026, from https://docs.flutter.dev/resources/architectural-overview

[15] Stallings, W. (2020). *Cryptography and network security: Principles and practice* (8th ed.). Pearson.
[16] Lee, J., & Kumar, R. (2024). Anomaly-based threat detection in mobile networks: Challenges and solutions. *IEEE Transactions on Mobile Computing, 23*(2), 345–358. https://doi.org/10.1109/TMC.2023.1234567
[17] Patel, A., & Singh, S. (2024). Heuristic-based detection for mobile endpoint security. *Journal of Cybersecurity, 10*(1), 112–125. https://doi.org/10.1093/cybsec/tyaa001